**Area between trajectories: Insights into optimal group selection and trajectory heterogeneity in group-based trajectory modeling**

Yi-Chen Hsiao[1,2,*], Chun-Yuan Chen[3,4,*], & Mei-Fen Tang[5,6,7]

[1] Institute of Health and Welfare Policy, School of Medicine, National Yang-Ming University, Taipei, Taiwan

[2] Big Data Analytics Program, Georgian College, Barrie, Ontario, Canada

[3] Institute of Health Policy and Management, College of Public Health, National Taiwan University, Taipei, Taiwan

[4] Analytics for Business Decision Making Program, Mohawk College, Hamilton, Ontario, Canada

[5] Department of Nursing, Wan Fang Hospital, Taipei Medical University, Taipei, Taiwan

[6] Research Center in Nursing Clinical Practice, Wan Fang Hospital, Taipei Medical University, Taipei, Taiwan

[7] School of Nursing, College of Nursing, Taipei Medical University, Taipei, Taiwan

*The first two authors contributed equally to this study.

**Corresponding Author**
Chun-Yuan Chen, PhD
Institute of Health Policy and Management
College of Public Health, National Taiwan University
Taiwan
Email: d98845002@ntu.edu.tw

**Abstract**

Group-based trajectory modeling (GBTM) is commonly used to identify longitudinal patterns in health outcomes among older adults, with determining the optimal number of groups being a crucial step. While statistically grounded criteria are primarily relied upon, clinical relevance is gradually emphasized in medicine to ensure that the identified trajectory heterogeneity appropriately reflects changes in a disease or symptom over time. However, such considerations are often judged through visual comparisons, without concrete approaches for their application. To address this, the Area Between Trajectories (ABTs) was introduced as insights for quantifying trajectory group differences. Using a simulated sleep quality dataset, GBTM was applied to build and compare models. Subsequently, ABTs was demonstrated to show how it works, while also highlighting its limitations and potential applications.

**Keywords**

Hsiao, Chen, Tang

***What this paper adds***

- The Area Between Trajectories (ABTs) was first introduced within group-based trajectory modeling (GBTM) as a direct, quantitative measure for comparing trajectory shapes beyond visual inspection.
- ABTs supports the selection of the optimal number of trajectory groups by providing a concrete benchmark for assessing trajectory heterogeneity and enabling potential linkage to clinical relevance and interpretability.

***Applications of study findings***

- ABTs offers potential extensions, such as assessing how far an individual's trajectory deviates from a target trajectory, investigating interval-specific risk factors, and serving as a measure for evaluating the effectiveness of an intervention.
- This brief report presents preliminary work on the potential of ABTs in GBTM. Future analyses using real-world data will aim to validate the concept and expand upon these initial observations.

**Introduction**

Group-based trajectory modeling (GBTM) has been applied to explore and classify longitudinal patterns of health outcomes in older adults, such as loneliness (Liu et al., 2021) and functional decline (Egbujie, Turcotte, Heckman, & Hirdes, 2024). Identifying distinct trajectories of a symptom or disease is valuable for guiding prevention and intervention when they reflect differences in risk factors, clinical progression, or treatment responses (Murray, Eisner, Nagin, & Ribeaud, 2022). For example, Egbujie et al. (2024) identified four functional decline trajectory groups (e.g., slow progressive decline, catastrophic decline). Members of these groups differed significantly in various individual-level factors, offering insights to develop personalized healthcare plans. In GBTM analyses, determining the optimal number of trajectory groups is a vital step, typically guided by several statistically grounded criteria.

Commonly used criteria include the Bayesian information criterion (BIC), the average posterior probability of assignment (APPA), and ensuring that the smallest group makes up at least 5% of the total sample (Egbujie et al., 2024). Although practices may vary across studies, determining the optimal number of groups typically involves key moves: (1) selecting criteria; (2) building and comparing models with an increasing number of groups, usually starting from two and stopping when the 5% minimum size rule is violated; and (3) deciding on the well-suited model based on a combined evaluation of the selected criteria. However, in medicine, it may not be advisable to rely solely on them for determining trajectory heterogeneity and classification, as this may sometimes construct trajectories that deviate from medical theories or clinical observations due to factors influencing the analyses (e.g., sample size).

Because of this, some studies have additionally considered clinical relevance in the decision-making process, ensuring that the resulting longitudinal patterns appropriately reflect the condition of interest. For instance, "The trajectory groups were also assessed for clinical utility…" (Dupre et al., 2025) and "the number of groups was determined based on clinical interpretability…" (Fukasawa et al., 2024). In other words, selecting the final model is ideally not reduced to a purely statistical side; clinical relevance considerations remain essential. Nevertheless, the ways these considerations are put into practice have not been detailed, leaving little guidance to follow. Very few studies mention it, but often vaguely. For example, Fukasawa et al. (2024) reported that models with more than four groups did not offer additional clinical relevance, without providing further explanation, suggesting that such judgments may largely rely on subjective visual comparisons in trajectory plots.

To our knowledge, visual comparisons remain the primary approach to assess the clinical relevance of different group solutions. While convenient, this approach has several limitations: (1) it is an imprecise way to assess differences between trajectories and can be easily influenced by plot arrangements, such as axis scaling; (2) these differences may appear visually subtle, yet they could carry important clinical significance; and (3) benchmarking one trajectory against another is challenging and often makes it difficult to pinpoint where (e.g., at specific intervals) meaningful differences lie in terms of clinical relevance. Therefore, we introduced a new approach—the Area Between Trajectories (ABTs)—to concretize and measure differences between trajectories. Area calculation is not brand new and has been widely applied in spatial analysis (Redecker et al., 2020). We believe it has potential to aid in determining the optimal number of groups in GBTM. While not intended to replace existing criteria, the ABTs serves as a supplementary role to enhance interpretability. As this brief report is set to be preliminary, more applications will be explored in future studies.

**Methods**

*Dataset and Variables*

A simulated dataset of 1,000 individuals was generated, including a unique identifier and an artificially generated total score (0–21) based on a seven-item Pittsburgh Sleep Quality Index (PSQI; Zitser et al., 2022) at each measurement point (week 0, week 2, …week 16). A function was customized to assign each individual to one of several predefined sleep quality trajectory groups and to simulate changes in PSQI scores across time, incorporating random variations to reflect real-world variability.

*Conceptual Framework*

The ABTs was defined as the area between two trajectories represented as lines or curves plotted in a coordinate system, where the x-axis denotes measurement time points and the y-axis represents the score range of a given scale. To prevent confusion with trajectory groups identified through GBTM, we refer to individual-level longitudinal patterns as individual trajectories. ABTs can be feasibly calculated over the entire observation period (Figure 1, Panel A), or within a specific time interval (Panel B). Also, ABTs can be applied to quantify the area between an individual trajectory and its assigned group (i.e., the representative trajectory of the membership in the group; Panels C and D).

[Figure 1]

*Data Analysis*

This exercise followed common GBTM practices, and all analyses were conducted using R version 4.5.0 (R Core Team, 2025). First, model comparison was based on three conventional criteria: the 5% minimum size rule, BIC (lower is better), and APPA (> 0.70). Second, the highest polynomial degree was only set to three to capture complex longitudinal patterns while minimizing overfitting. Third, GBTM was performed to build models with varying numbers of groups ('gbmt' package). A few models were expected to stand out as candidate models, which then were used for illustrations. Fourth, the trapz() function was used to approximate *ABTs* across all intervals, with each interval divided into 1,000 discrete segments ('pracma' package). As the mathematical details of area calculation are well-documented elsewhere, they are not elaborated upon here. Finally, the distribution of all interval-specific ABTs across group comparisons was explored.

**Results and Illustrations**

*Evaluating Models Using Conventional Criteria*

Models with six or more groups were excluded from consideration because the smallest group size was less than 5% of the total sample. Therefore, only solutions with 2-5 groups were compared. If only BIC and APPA were considered (Table 1), the five-group model would typically be preferred. However, its trajectory heterogeneity profile was similar to that of the four-group solution (Figure 2). The main difference was that the individuals with consistently lower scores across all time points were classified as a single group (Group #1; good and stable sleep quality) in the four-group model but appeared to be separated into two groups (Groups #1 and #5) in the five-group model. This raises a question: is it necessary to distinguish between the two groups?

[Table 1]

[Figure 2]

*ABTs and Their Distributions*

As an illustrative example, the interval-specific ABTs between Group #5 and Group #1 in the five-group solution were estimated. The values ranged from 4.08 to 6.02, yielding a total of 41.54 (Figure 3, Panel A). This approach provides a quantitative benchmark for assessing the degree of distinction between trajectories, as opposed to visual comparisons. In addition, ABTs can serve as an individual-level reference for intervention, indicating the extent of 'effort' required for a person to align with a target trajectory (e.g., the consistently good sleep pattern). For example, the total area between Individual #51 and Group #1 was 70.83 (Panel B).

[Figure 3]

Further, another question remains: how meaningful is the degree of distinction between Group #5 and Group #1? This question can be initially responded by plotting the distributions of interval-specific areas between all pairs of groups. In Figure 4, several types of distribution were observed, offering potential insights: (1) when the distribution is positioned further to the left on the x-axis, it suggests that the two trajectories are relatively closer compared to other group comparisons; conversely, distributions positioned further to the right indicate greater separation; and (2) a concentrated and narrow distribution suggests that the two trajectories are relatively parallel over time, whereas a flatter one may indicate that the trajectories diverge, converge, or fluctuate irregularly with time. In our case, the distribution for Group #5 versus Group #1 was the most left-shifted, concentrated, and narrow, indicating that the two trajectories were relatively close and parallel over time.

*[Figure 4]*

**Discussion**

In GBTM, a critical step involves determining the optimal number of trajectory groups to adequately capture the heterogeneity in the outcome of interest. In medicine, identifying distinct trajectories that carry reasonable and interpretable clinical meaning holds more weight than simply selecting the number of groups based solely on statistically grounded criteria (Dupre et al., 2025; Fukasawa et al., 2024). However, how such considerations are applied in practice remains unclear. Heavily relying on statistically grounded criteria may lead to redundant groups that exhibit highly similar levels and shapes but lack meaningful clinical distinctions (Dupre et al., 2025). Further, in certain cases, some criteria (e.g., BIC) tend to continue improving as more trajectory groups are added to the model (Jarry et al., 2025).

In conventional approaches, several challenges may arise. First, the statistically preferred number of groups does not necessarily align with clinically meaningful distinctions between groups. Second, direct, quantitative measures to assess differences between trajectories are often lacking, leaving judgments largely based on visual comparisons. Third, no benchmarks support the consideration of clinical relevance. Lastly, it is not uncommon for multiple models to show comparable statistical performance, leading to a dilemma in model selection. Therefore, ABTs was introduced with the belief that it holds potential to enhance the decision-making process by offering a benchmark and serving as a possible means of linking model determination to clinical relevance, especially in situations where different models exhibit highly similar trajectory heterogeneity and comparable performance on statistically grounded criteria.

In our exercise using simulated sleep quality data, the five-group solution was favored by conventional criteria. However, two of the groups appeared highly similar in their trajectory levels and shapes. With respect to level, both trajectories consistently hovered around a PSQI score of five or lower throughout the observation period, aligning with the threshold for good sleep (Zitser et al., 2022). Their similarity was also supported by ABTs values, which fell within a narrow range of 4 to 6. In terms of slope, both trajectories kept nearly parallel over time, as indicated by the ABTs, which showed relatively minor fluctuations across intervals compared to other group comparisons. These observations suggest that the four-group solution may be sufficient. Similar discussions appeared in previous studies. For example, separating a large 'low' conduct-disorder trajectory group into 'low' and 'no' may offer limited benefits (Nagin, Jones, & Elmer, 2024). Please note that the sleep quality example serves solely to highlight the potential of ABTs in informing group selection and offering an alternative to visual comparisons, rather than to establish clinical distinctiveness between the groups.

Several potential extensions of ABTs are envisioned. ABTs can be also applied to assess how far an individual's trajectory deviates from a target trajectory. This can help estimate the 'effort' required for an individual to catch up with the goal and identify specific time intervals that may demand greater intervention. Additionally, ABTs introduces the possibility of investigating interval-specific risk factors, offering a more granular understanding. This also opens the window to linking differences between trajectories to clinical relevance through clinical indicators. Despite the strengths and potentials, several limitations should be noted. First, the ABTs was not tailored to provide absolute values but rather serves as relative references for comparing trajectories. Their interpretation may be influenced by factors such as measurement time units; therefore, caution is warranted when interpreting ABTs, as their magnitude, to some degree, depends on the coordinate system. Second, ABTs in nature does not convey clinical relevance; rather, additional work is required to dive into how they can be meaningfully linked to clinical indicators of interest. Third, this insight is not intended to perform statistical significance testing but rather to provide an interpretable quantitative reference.

## Conclusion

This study introduced the ABTs as insights into optimal group selection and trajectory heterogeneity in GBTM. The ABTs provides a measure of differences between groups and serve as an alternative to visual comparisons, with the potential to act as a potential bridge between model selection and clinical relevance. While it has limitations and is not intended to replace existing criteria, ABTs is expected to serve a supplementary role.

## Acknowledgments

YCH, CYC, and MFT contributed to the conceptualization and design of this work. YCH and CYC conducted the literature review, data simulation, formal analysis, and wrote the original draft. MFT provided feedback on the manuscript. All authors read and approved the final version.

## Declaration of Conflicting Interests

The author(s) declared no conflicts of interest related to the research, authorship, or publication of this article.

## Funding

This work received no specific funding from public, commercial, or not-for-profit agencies.

**Ethical Statement**

This work was exempt from ethical review because it used only simulated data and involved no human or animal participants.

## Figures and Tables

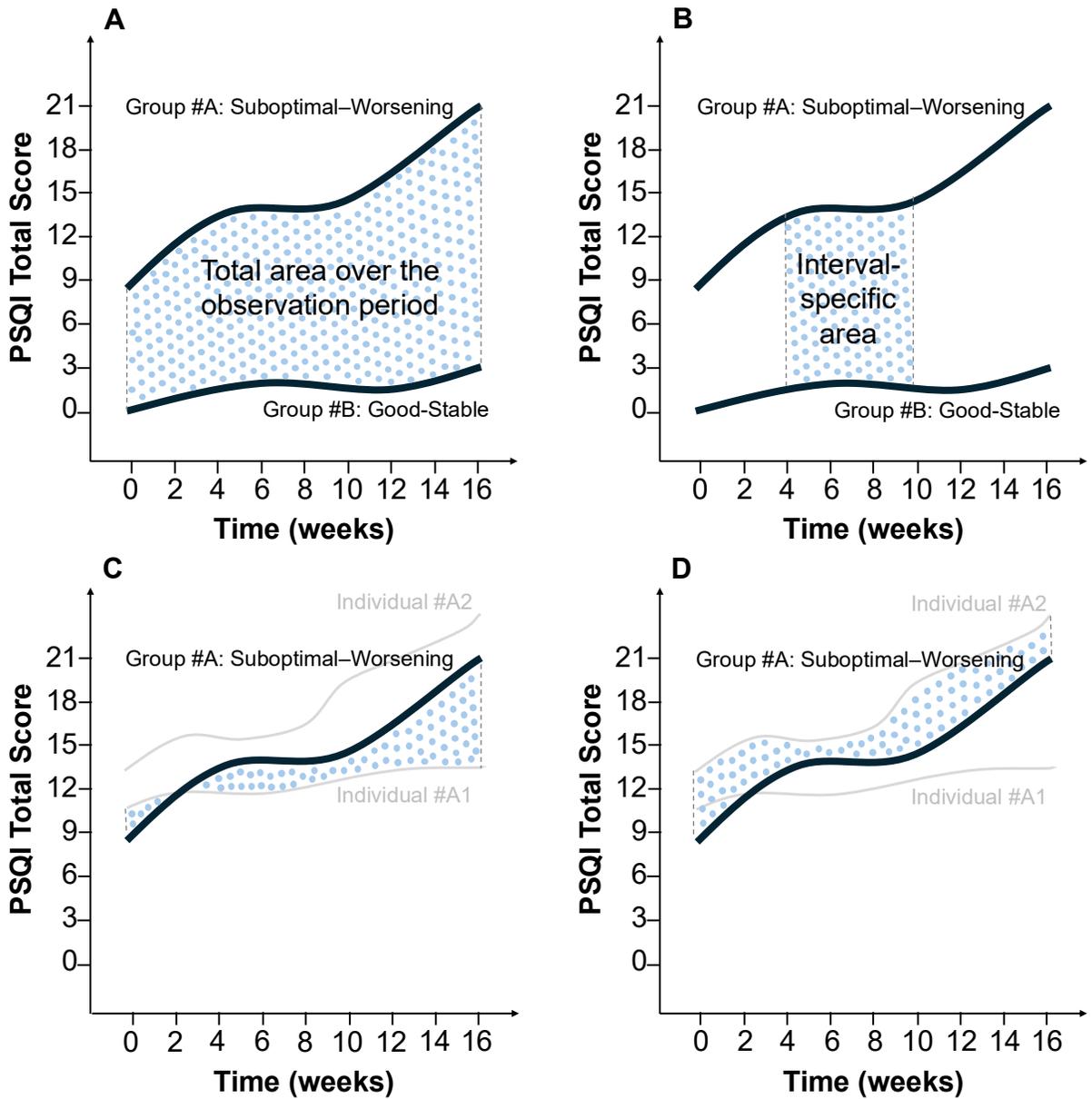

**Figure 1.** Illustration of area-based comparisons between trajectories.
A. Total area (dots) between two trajectory groups across the full observation period. B. Area between two trajectory groups within a specific time interval. C and D. Area between an individual trajectory and its assigned trajectory group.

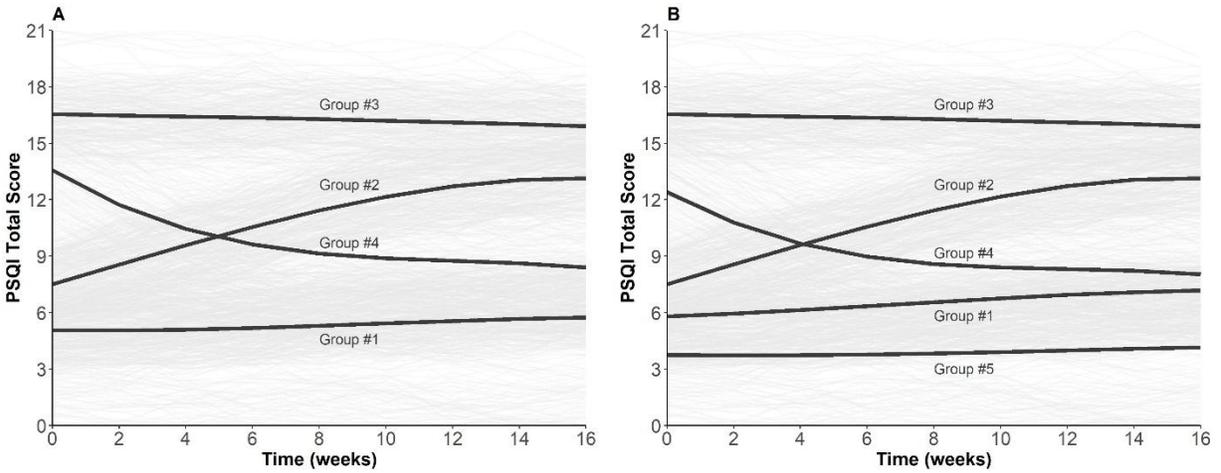

**Figure 2.** Levels and shapes of sleep quality trajectory groups.

A: Four-group solution. B: Five-group solution.

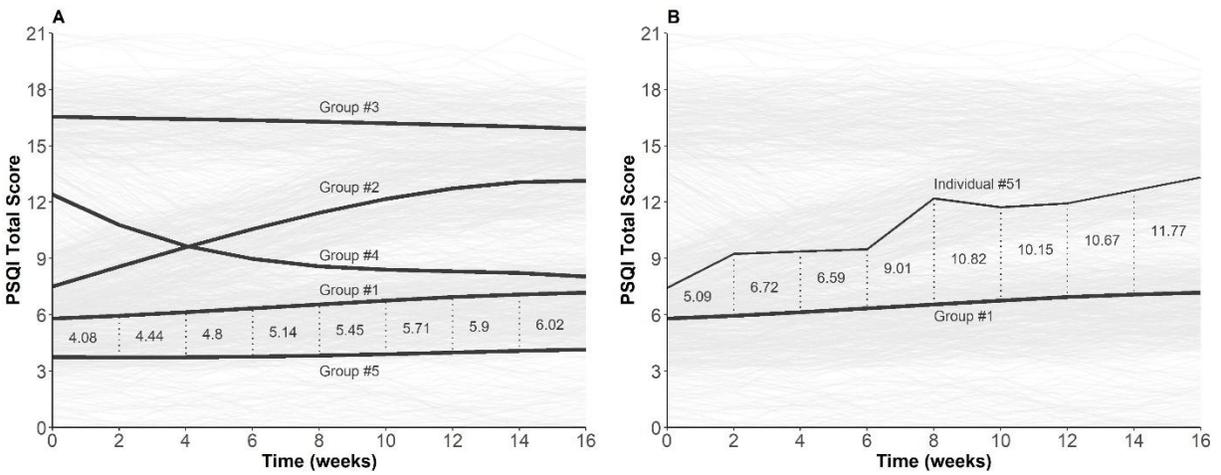

**Figure 3.** Interval-specific areas between trajectories.

A: Group #5 to Group #1. B: Individual #51 to Group #1.

Hsiao, Chen, Tang

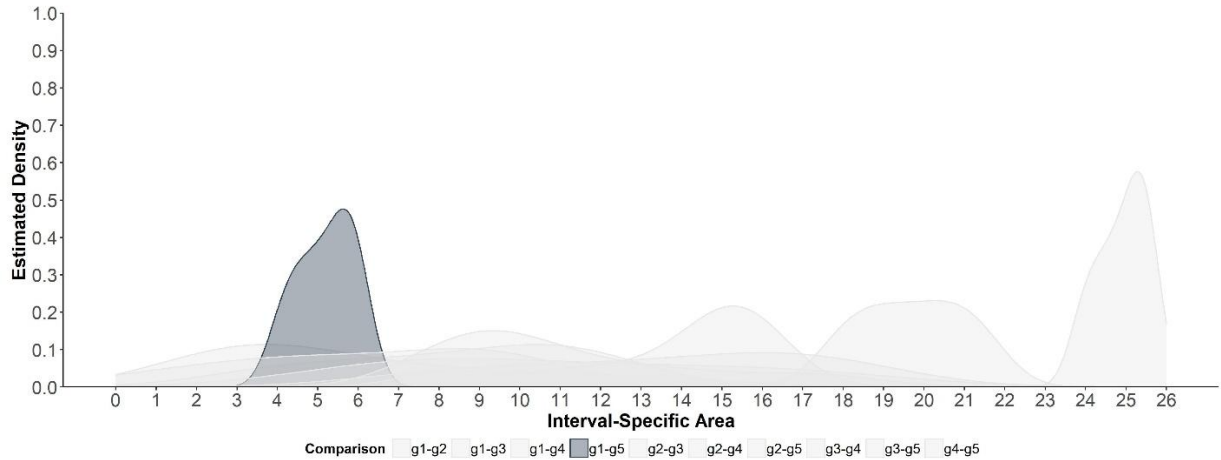

**Figure 4.** Distribution of interval-specific areas between all pairwise group comparisons.

**Table 1.** Model fit indices across models with varying numbers of trajectory groups.

| Model with number of groups specified[a] | The smallest group % | BIC[b] | APPA[c] |
|---|---|---|---|
| 2 | 44.3 | 44549 | 0.997 |
| 3 | 23.7 | 38870 | 0.995 |
| 4 | 8.5 | 37030 | 0.995 |
| 5 | 11.4 | 34337 | 0.990 |

[a]The number of groups specified began at two. Models with six or more groups were not considered, as they violated the 5% minimum group size criterion.

[b]The sample size-adjusted BIC was also examined and showed a consistent pattern with the traditional BIC reported in the table.

[c]It is important to note that our data were simulated to demonstrate ABTs, so APPA values in real-world datasets are generally expected to be lower.